\documentclass[12pt]{article}
\usepackage[margin=25mm]{geometry}
\usepackage{amsmath, amssymb, amsthm}
\usepackage{graphicx}
\usepackage[authoryear]{natbib}
\usepackage{url} 
\usepackage{array}
\usepackage[normalem]{ulem} 
\usepackage{xcolor}
\usepackage{multirow}
\usepackage{IEEEtrantools}
\usepackage{mathtools}
\usepackage{comment}
\usepackage{bbm}
\usepackage{multirow}
\usepackage{diagbox}
\usepackage{booktabs}
\usepackage{bigints}
\usepackage{bm}
\usepackage[colorlinks, citecolor=blue, urlcolor=blue]{hyperref}
\usepackage{algorithm,algorithmic}
\usepackage{siunitx}
\usepackage{xr}

\newcommand{\I}{\mathcal{I}}

\newcommand{\D}{\mathcal{D}}

\newcommand{\T}{\mathcal{T}}

\newcommand{\E}{\mathbb{E}}

\newcommand{\N}{\mathbb{N}}

\newcommand{\Cov}{\mathbb{C}\text{ov}}
\def\bU{\mathbf{U}}
\def\bB{\mathbf{B}}
\def\bM{\mathbf{M}}
\def\bE{\mathbf{E}}
\def\bG{\mathbf{G}}
\def\bI{\mathbf{I}}
\newcommand{\eps}{\varepsilon}

\newcommand{\what}[1]{\widehat{#1}}
\newcommand{\innerproduct}[2]{\left\langle #1, #2 \right\rangle}
\DeclarePairedDelimiterX{\norm}[1]{\lVert}{\rVert}{#1}
\DeclarePairedDelimiterX{\abs}[1]{\lvert}{\rvert}{#1}

\newcommand{\brac}[1]{\left \{ #1 \right \}}
\DeclarePairedDelimiterX{\sbrac}[1]{[}{]}{#1}
\DeclarePairedDelimiterX{\pbrac}[1]{(}{)}{#1}

\newcommand{\amin}{\arg\min}

\def\hphi{\hat{\phi}}
\def\hlambda{\hat{\lambda}}

\def\t{\mathbf{t}}

\numberwithin{thm}{section}

\numberwithin{cor}{section}

\numberwithin{prop}{section}
\newcommand{\proofpart}[2]{%
  \par
  \addvspace{\medskipamount}%
  \noindent \textbf{Part #1:} #2\par\nobreak
  \addvspace{\smallskipamount}%
  \@afterheading
}

\numberwithin{remark}{section}

\providecommand{\keywords}[1]
{
  \small	
  \textbf{\textit{Keywords---}} #1
}
\def\spacingset#1{\renewcommand{\baselinestretch}%
{#1}\small\normalsize} \spacingset{1}

\title{\bf Order Determination for Functional Data}
\author{Chi Zhang, Peijun Sang, and Yingli Qin\\
        \small Department of Statistics and Actuarial Science, University of Waterloo \\
}
\date{}
\begin{document}
\maketitle
\begin{abstract}
Dimension reduction is often necessary in functional data analysis, with functional principal component analysis being one of the most widely used techniques. A key challenge in applying these methods is determining the number of eigen-pairs to retain, a problem known as order determination. When a covariance function admits a finite representation, the challenge becomes estimating the rank of the associated covariance operator. While this problem is straightforward when the full trajectories of functional data are available, in practice, functional data are typically collected discretely and are subject to measurement error contamination. This contamination introduces a ridge to the empirical covariance function, which obscures the true rank of the covariance operator. We propose a novel procedure to identify the true rank of the covariance operator by leveraging the information of eigenvalues and eigenfunctions. By incorporating the nonparametric nature of functional data through smoothing techniques, the method is applicable to functional data collected at random, subject-specific points. Extensive simulation studies demonstrate the excellent performance of our approach across a wide range of settings, outperforming commonly used information-criterion-based methods and maintaining effectiveness even in high-noise scenarios. We further illustrate our method with two real-world data examples.
\end{abstract}
\vfill

\newpage
\spacingset{1.8}
\section{INTRODUCTION}\label{sec:intro}
Functional data analysis (FDA) provides a framework for analysing functional data that vary continuously over a domain, such as time or space. The intrinsically infinite-dimensional nature of functional data necessitates the use of dimension reduction techniques, which transform infinite-dimensional random functions into finite-dimensional random vectors, in many applications. This transformation allows for subsequent analysis using tools from multivariate analysis. Moreover, dimension reduction is commonly adopted as a way of regularization when inverting the covariance operator of functional data, as required in functional linear regression \citep{hall_methodology_2007,
Zhou_Yao_Zhang_2023} and functional generalized linear models \citep{dou_estimation_2012}. Generally, without proper regularization, the inverse of the covariance operator is unbounded, which renders it difficult to fit these models. Among various dimension reduction methods, functional principal component analysis (FPCA) has garnered significant attention due to its ability of using a parsimonious subspace to explain the most relevant variation around a mean function in a data-adaptive manner \citep[e.g.,][]{Yao_Muller_Wang_2005_FPCA, hall_properties_2006, hall2006properties}.

A subtlety in FPCA is the selection of the number of functional principal components (FPCs) to retain. A common observation is that higher-order FPCs often exhibit significant variations, making their interpretation difficult. This leads to the pragmatic assumption that the covariance operator has finite rank $d$, treating higher-order terms as noise \citep{li_selecting_2013}. In essence, this transforms the order determination problem to estimating the rank of the covariance operator. When functional data are fully observed without any measurement error, the rank can be estimated in a straightforward manner. This follows from the fact that the estimated covariance operator is a linear combination of observed trajectories, which themselves can be expressed as linear combinations of eigenfunctions associated with non-zero eigenvalues via the Karhunen–Lo\`{e}ve expansion. However, in practice, functional data are often observed discretely and contaminated by measurement errors. This introduces a confounding issue: the true rank of the covariance operator becomes obscured by the presence of noise, which effectively adds a ridge to the true covariance function.

Estimating the rank of the covariance operator from contaminated functional data can be approached through heuristic methods such as scree plots and the fraction of variance explained (FVE), which normally require a subjective pre-specified threshold. A more deliberate approach is to separate the effect of measurement errors from the true covariance function. \cite{Hall_Vial_2006} proposed a ``low-noise" model, assuming that the noise variance diminishes as the sample size increases. However, this method still requires a subjective threshold as a stopping criterion. \cite{charkaborty_testing_2022} proposed an alternative approach by approximating the infinite-dimensional functional space with a finite-dimensional matrix space. They argued that the corruption of the diagonal entries of the covariance matrix by measurement errors does not affect the rank estimation. Thus, they disregarded the diagonal entries of the sample covariance matrix and subsequently filled them via matrix completion based on a modified Frobenius distance, yielding a matrix with the same rank as the discretized covariance function.

Another approach to removing the impact of measurement errors entails smoothing techniques that leverage the continuity of the covariance function. Once a smoothed covariance function is obtained, the problem can be framed as a model selection problem, allowing the use of information criterion (IC)-based methods, such as the Akaike information criterion (AIC) \citep{Yao_Muller_Wang_2005_FPCA} or the Bayesian information criterion (BIC) \citep{fdapace}. However, a direct application of classical IC-based techniques to functional data tends to favour selecting an excessive number of FPCs, or equivalently, overestimating the rank. This issue may arise from the nonparametric nature of the data, where each FPC comprises both a variance parameter and a nonparametric function \citep{li_selecting_2013}. To address this, \cite{li_selecting_2013} introduced modified penalty terms in place of those used in AIC and BIC. The penalty term for the adjusted BIC method depends on eigenvalues approaching zero, which pose a challenge, since small eigenvalue estimates can be unreliable in practice. The modified penalty term for AIC is derived under a Gaussian assumption for densely observed functional data. Furthermore, AIC-based methods require that the true model is within the set of candidate models, disregarding potential estimation bias \citep{hurvich_smoothing_1998, li_selecting_2013}. This assumption, however, is fundamentally flawed when nonparametric smoothing is applied to estimate the mean and covariance functions, as bias is inherently introduced by those smoothing methods. 

When estimating the rank of the covariance operator of functional data, many existing methods often rely solely on estimated eigenvalues (\citealt{Hall_Vial_2006}, the modified BIC in \citealt{li_selecting_2013}, and \citealt{charkaborty_testing_2022}), without incorporating information from estimated eigenfunctions.
In this paper, inspired by the work of \cite{Luo_Li_2016} on determining the number of principal components for multivariate data, we propose a novel procedure for estimating the rank of the covariance operator by integrating information from both estimated eigenvalues and eigenfunctions. To the best of our knowledge, no existing FDA method integrates information from both estimated eigenvalues and eigenfunctions for rank estimation. Furthermore, unlike AIC-type methods (e.g., \citealt{Yao_Muller_Wang_2005_FPCA} and the modified AIC in \citealt{li_selecting_2013}), our method does not require a distribution assumption to estimate the FPC scores. The key observation is that the variability of the estimated eigenfunctions increases sharply when their index exceeds the true rank $d$, while the corresponding estimated eigenvalues exhibit a steep drop.

Unlike the multivariate setting studied by \cite{Luo_Li_2016}, we need to account for the intrinsic infinite-dimensional nature of functional data and sparse observations that are contaminated by measurement errors. In particular, we estimate the mean and covariance functions by applying a local linear smoother to aggregated observations. Instead of using the entire dataset to estimate eigenfunctions, we partition the subjects into two disjoint subsets and estimate their eigenfunctions separately to assess the variability of these eigenfunction estimates. Combining this variability assessment with eigenvalues estimated from the complete dataset, we develop the Functional Ladle Estimator (FLE) to determine the rank of the covariance operator. The numerical studies showcase excellent performance of our method under various simulation settings, whereas IC-based methods are sensitive to the choice of simulation settings. When applying our method as well as IC-based methods to two real-world applications, FLE also displays great advantages in estimating the order. 

The remainder of this paper is organized as follows. Section \ref{sec:FPCA} introduces the data generation process and provides an overview of the FPCA estimation procedures. Section \ref{sec:function_ladle_estimator} provides a detailed description of FLE. In Section \ref{sec:simulation} we conduct extensive simulation studies to compare the performance of our method with that of some alternative methods in various settings, and in Section \ref{sec:real_data} we apply FLE and some alternatives to two real-world datasets: bike-sharing and air pollution data. We conclude with a discussion in Section \ref{sec:conc}.
\section{FUNCTIONAL PRINCIPAL COMPONENT ANALYSIS}\label{sec:FPCA}
\subsection{Data Structure and Model Assumptions}
Let $X(t)$ be a continuous and square-integrable stochastic process defined on a compact interval $\T=[0,1]$, with mean function $\mu(t)$ and  covariance function $G(s,t) = \E \{X(s)-\mu(s)\}\{X(t) - \mu(t)\}$. 
Under the continuity assumption on $X$, this covariance function defines an operator from $L^2([0, 1])$ to $L^2([0, 1])$: $(\bG f)(s) = \int_0^1 G(s, t) f(t) dt$ for any $f \in L^2([0, 1])$. Furthermore, 
the covariance function can be represented as
\begin{equation}\label{eq:Mercer}
    G(s,t) = \sum_{\nu=1}^{\infty}\lambda_\nu\phi_\nu(s)\phi_\nu(t), \quad t,s \in \T,
\end{equation}
where $(\lambda_\nu, \phi_\nu)$ is the $\nu$th eigenvalue-eigenfunction pair of $\bG$ satisfying $\bG \phi_{\nu} = \lambda_{\nu} \phi_{\nu}$ and these $\phi_{\nu}$'s form an orthonormal basis of $L^2([0, 1]$. 
Without loss of generality, we assume that $\lambda_1 > \lambda_2 > \ldots > 0$. 

As noted in Section \ref{sec:intro}, we assume that the covariance operator has finite rank $d$, implying $\lambda_\nu = 0$ for all $\nu > d$ in \eqref{eq:Mercer}. We say the dimensionality of $X$ is $d$ under this assumption. 
Consequently, the Karhunen–Lo\`{e}ve expansion of $X(t)$ reduces to
\begin{equation}\label{eq:KL_expansion_truncated}
    X(t) = \mu(t) + \sum_{\nu=1}^{d}\xi_\nu\phi_{\nu}(t), ~~t \in \T,
\end{equation}
where $\xi_\nu = \int_{\T} \{X(t)-\mu(t)\}\phi(t) dt,~\nu=1, 2, \ldots, d$ are uncorrelated zero-mean random variables with variance $\lambda_\nu$.
Given $n$ i.i.d. sample paths of $X$, we assume that the responses $Y_{ij}$ are observed at discrete time points $T_{ij}$ from $X_i$, subject to additive measurement errors:
\begin{equation}\label{eq:model}
    Y_{ij} = X_{i}(T_{ij}) + \eps_{ij}, \quad i=1, 2, \ldots, n, j=1, 2, \ldots, N_i.
\end{equation}
Here, $N_i$'s can be random or fixed, and $T_{ij}$'s are subject-specific (potentially random) observation times in $\T$. Measurement errors $\eps_{ij}$ are independent random variables with mean zero and variance $\sigma^2_{\eps}$. Moreover, $N$, $T$, $\eps$ and $X$ are independent. Model \eqref{eq:model} is widely adopted in modelling longitudinal observations under the framework of functional data; see \cite{Yao_Muller_Wang_2005_FPCA}, \cite{li2010uniform}, \cite{Zhang_Wang_2016} and references therein.  
For convenience, we define $R(T_{ij}, T_{il}) = \Cov(Y_{ij}, Y_{il}) = G(T_{ij}, T_{il}) + \sigma_\eps^2 \delta_{jl}$, where $\delta_{jl} = 1$ if $j = l$, and $0$ otherwise.

\subsection{Estimation of the Model Components}\label{sec:model_compoenent_estimation}
The proposed method is based on the estimation of the eigenpair $(\lambda_{\nu}, \phi_{\nu})$, which are normally obtained by performing the spectral decomposition on the discretization of the smoothed covariance function \citep{rice_estimating_1991}. The selection of an appropriate method to estimate the mean function $\mu(\cdot)$ and the covariance function $G(\cdot, \cdot)$ depends on the sampling rate and sampling scheme of $X$; see \cite{Cai_Yuan_2011} and \cite{Zhang_Wang_2016} for a more detailed discussion. Here we estimate $\mu(\cdot)$ and  $G(\cdot, \cdot)$ by pooling observations across subjects, following the approach of \cite{Yao_Muller_Wang_2005_FPCA}. In particular, we employs a local linear smoother to estimate $\mu(\cdot)$, where $\hat{\mu}(t) = \hat{a}_0$ is given by
\begin{equation*}\label{eq:local_linear_mu_est}
    (\hat{a}_0, \hat{a}_1) = \underset{a_0, a_1}{\amin} \sum_{i=1}^{n}\sum_{j=1}^{N_i}K_1\left(\frac{T_{ij} - t}{h_\mu}\right)\brac{Y_{ij} - a_0 - a_1(T_{ij} - t)}^2,
\end{equation*}
where $K_1(\cdot)$ is a symmetric kernel function, and $h_\mu$ denotes the bandwidth for the estimation of $\mu(\cdot)$. 
We adopt a similar method to estimate $G(\cdot,\cdot)$. Specifically, $\hat{b}_0 = \widehat{G}(s, t)$ is determined by using a local linear surface smoothing technique as follows:
\begin{equation*}\label{eq:local_linear_cov_est}
\begin{aligned}
    (\hat{b}_0, \hat{b}_1, \hat{b}_2) = \underset{b_0, b_1, b_2}{\amin}\sum_{i=1}^{n}\sum_{1 \leq j\neq l \leq N_i}&\brac{\widehat{R}_{i}(T_{ij}, T_{il}) - b_0 - b_1(T_{ij} - t) - b_2(T_{il} - s)}^2\times\\
    &~K_2\left(\frac{T_{ij} - t}{h_G}, \frac{T_{il} - s}{h_G}\right),
\end{aligned}
\end{equation*}
where $K_2(\cdot, \cdot)$ is a symmetric bivariate kernel function, with $h_G$ being the bandwidth for estimating $G(\cdot, \cdot)$. The eigenpairs $\{\lambda_\nu, \phi_\nu(\cdot)\}_{\nu=1}^{L}$ can be estimated by performing an eigen-decomposition on the discretized estimated covariance function; see Chapter 8 of \cite{ramsayFDA2005} for more details. Let $L < n$ be a prespecified number that is larger than $d$, denoting the maximum index in the search range for $d$. In practice, $L$ can be chosen as the total number of non-negative eigenvalues of the discretized $\what{G}$ or determined by setting a sufficiently large threshold for the FVE, such as $99.99\%$ \citep{zhu2014structured}, i.e., 
$$
L = \min \left\{\nu: \frac{\sum_{j = 1}^{\nu} \hat{\lambda}_j}{\sum_{j \geq 1} \hat{\lambda}_j} \geq .9999\right\}. 
$$

\section{FUNCTIONAL LADLE ESTIMATOR}\label{sec:function_ladle_estimator}
To bypass the Gaussian assumption, which is needed in the modified AIC in \cite{li_selecting_2013}, and effectively leverage the variability pattern of the estimated eigenfunctions, we develop the following procedure to estimate the rank of the covariance operator $\bG$. As a preliminary step, we employ a sample-splitting strategy, randomly partitioning the entire dataset into two disjoint subsets, $\D_{1}$ and $\D_{2}$. Without loss of generality, we denote
\begin{gather*}
    \D_{1} = \brac{\left(Y_{ij}, T_{ij}\right) \mid i \leq n/2;~j\leq N_i;~i,j \in \N_{+}},\\
    \D_{2} = \brac{\left(Y_{ij}, T_{ij}\right) \mid n/2 <i \leq n;~j\leq N_i;~i,j \in \N_{+}}.
\end{gather*}
For $g=1, 2$, let $\widehat{G}_{g}$ and $\big\{\hlambda_{g,\nu}, \hphi_{g,\nu}(\cdot)\big\}_{\nu=1}^{L}$ denote the estimated covariance function and the corresponding eigenpairs derived from subset $\D_{g}$, using the techniques introduced in Section \ref{sec:model_compoenent_estimation}. The sampling splitting strategy enables us to quantify the variability in eigenfunction estimates, as detailed below. This strategy is also adopted in \cite{zhou_wei_yao_2024}.

\begin{figure}
    \centering
    \includegraphics[scale = 0.4]{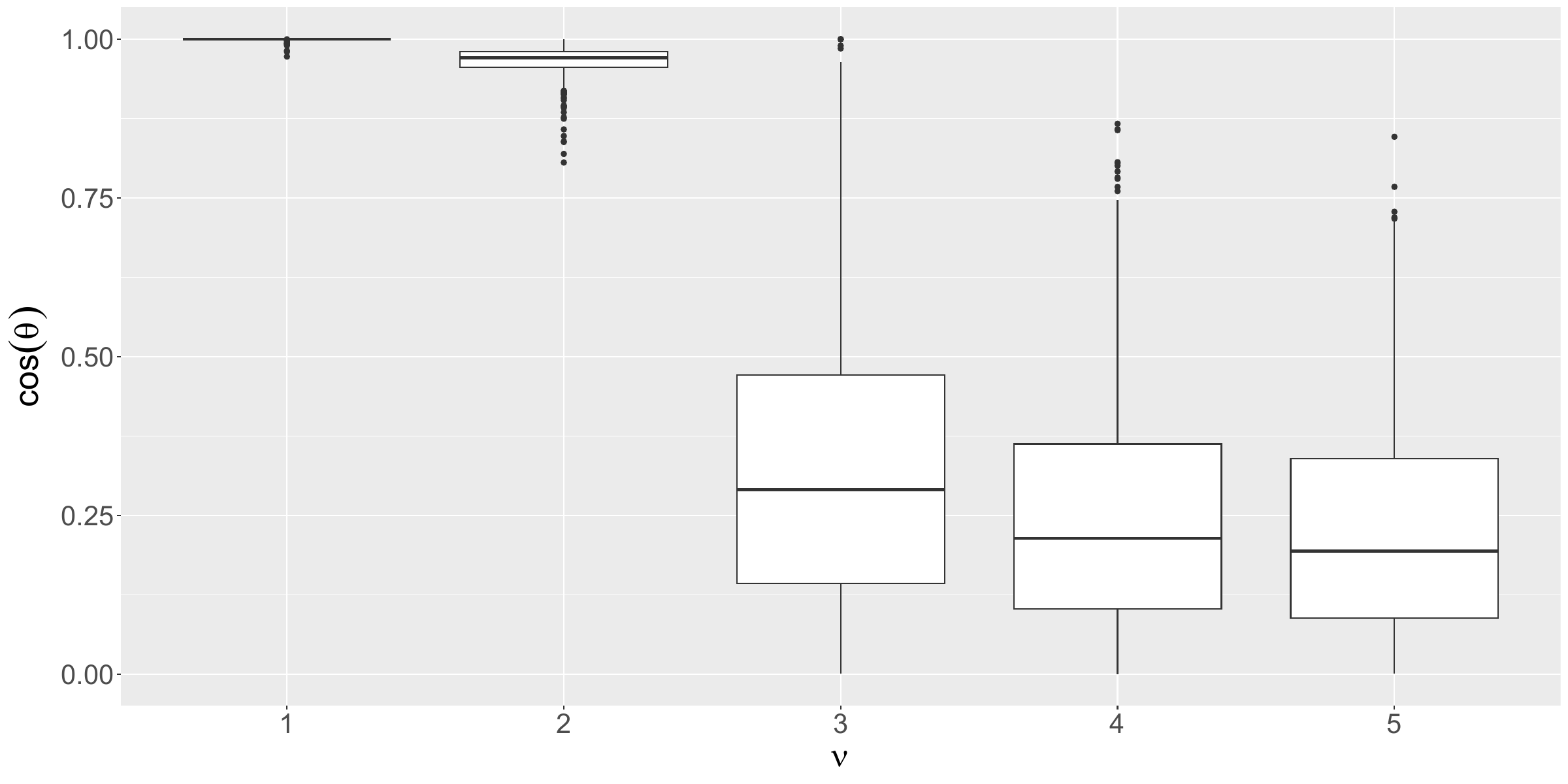}
    \caption{The boxplot for $\cos{\theta}:= \langle\hphi_{1,\nu},\hphi_{2,\nu}\rangle$ for $\nu = 1, 2, \ldots, 5$ over 1000 simulation runs, where the true rank is $2$.}
    \label{fig:angle_boxplot}
\end{figure}

Under mild conditions, for $\nu \leq d$,  the estimated eigenfunctions $\hphi_{g,\nu}$ converge in probability to $\phi_\nu$ in both the $L^2([0, 1])$ and the $L^{\infty}([0, 1])$ norms as the sample size increases for $g=1, 2$ \citep{zhou_wei_yao_2024}. The consistent estimation of $\phi_\nu$ implies that $\langle\hphi_{1,\nu},\hphi_{2,\nu}\rangle \approx 1$, where $\langle f, g \rangle = \int_0^1 f(t)g(t) dt$ for any $f, g \in L^2([0, 1])$. However, for $\nu > d$, $\hphi_{g,\nu}$ is no longer a consistent estimator for $\phi_{\nu}$ as $\lambda_{\nu} = 0$. It should be noted that $\phi_{\nu}$ is not well defined at the population level since $\lambda_{\nu}$ = 0 for $\nu > d$. However, at the sample level, we may still obtain $\hphi_{\nu}$ from the eigendecomposition of $\what{G}$ even for $\nu > d$. These estimates arise from the measurement errors in model \eqref{eq:model}, as well as the estimation errors introduced by local linear smoothing. Although $\what{G}$ consistently estimates $G$ under mild conditions \citep{zhou_wei_yao_2024}, the re-normalization step in the spectral decomposition on $\what{G}$ amplifies the variability of $\hphi_{\nu}$'s with index $\nu > d$. Therefore, $\langle\hphi_{1,\nu},\hphi_{2,\nu}\rangle$ could be much smaller than $1$ for $\nu > d$ with high probability. Figure \ref{fig:angle_boxplot} 
displays the box plots of $\langle\hphi_{1,\nu},\hphi_{2,\nu}\rangle$ across 1000 simulation runs for different values of $\nu$, where the true rank of $\bG$ is set to $2$. More details on the data generation process for Figure \ref{fig:angle_boxplot} are available in \ref{appendix:FLE_sec}.
Figure \ref{fig:angle_boxplot} justifies our assertion about the phase transition of $\langle\hphi_{1,\nu},\hphi_{2,\nu}\rangle$ from close to 1 to much smaller than 1.  
This observation motivates us to introduce a Gram matrix $\bU_\ell$ for $\ell \leq L$, which is defined as
\begin{equation*}
\bU_\ell = (u_{ij})_{\ell \times \ell}, \quad u_{ij} = \innerproduct{\hat{\phi}_{1,i}}{\hat{\phi}_{2,j}}, \quad\text{for } i, j \leq \ell. 
\end{equation*}
Then we define $f_{\bU}(\ell) = 1 - \abs{\det(\bU_\ell)}$ for $1 \leq \ell \leq L$ to quantify the variability of the space spanned by the first $\ell$ estimated eigenfunctions.

Intuitively, $\bU_\ell$ converges to the identity matrix in probability when $\ell \leq d$. In contrast, when $\ell > d$, let us partition the matrix $\bU_\ell$ into $4$ submatrices as follows:
\begin{equation}\label{eq:mat_partition}
\bU_\ell = \left[
\begin{array}{c|c}
\bM_1 & \bE_1 \\
\hline
\bE_2 & \bM_2
\end{array}
\right], 
\end{equation}
where $\bM_1$ is a $d \times d$ matrix and $\bM_2$ is an $(\ell-d) \times (\ell-d)$ matrix. By Corollary 5.1.5 in \cite{hsing_theoretical_2015} and Theorem 4.2 in \cite{Zhang_Wang_2016}, $u_{ij} \approx 0$ with high probability when $i > d, j \leq d$ or $i \leq d, j > d$, leading to $\det(\bU_\ell) \approx \det(\bM_2)$. Meanwhile, the diagonal entries of $\bM_2$ are much smaller than $1$, which implies a much smaller $\abs{\det(\bU_\ell)}$. 

To find $f_{\bU}(\ell)$, we evaluate $\hphi_{g, \nu}(t)$ at a fixed grid $\t = (t_1, t_2, \ldots, t_m)^{\top}$, where the grid points are equally spaced with size $\delta$, i.e., $\delta = t_{i+1} - t_{i}$ for $i=1, 2, \ldots, m-1$. Let $\widehat{\bB}_{g,\ell} = \brac{\hat{\phi}_{g,1}(\t), \hat{\phi}_{g,2}(\t), \dots, \hat{\phi}_{g,\ell}(\t)}$ for $g = 1, 2$. Provided that $\delta$ is sufficiently small, each entry of $\delta\widehat{\bB}_{1,\ell}^{\top}\widehat{\bB}_{2,\ell}$ is a Riemann sum of the corresponding entry in $\bU_\ell$, and thus, $1 - \abs{\det({\delta\widehat{\bB}_{1,\ell}^{\top}\widehat{\bB}_{g,\ell}})} := \hat{f}_{\bU}(\ell) $ serves as an approximation of $f_{\bU}(\ell)$, quantifying the discrepancy between the columns space of $\widehat{\bB}_{1,\ell}$ and $\widehat{\bB}_{2,\ell}$. When $\ell \leq d$, the eigenfunctions $\phi_\nu$ are consistently estimated by $\hat{\phi}_{g,\nu}$'s for all $\nu \leq \ell$ and $g=1,2$, leading to a minor discrepancy. However, for $\ell > d$, the estimates of $\phi_{\nu}$ become increasingly dominated by noise for $\nu > d$, leading to a significant increase in discrepancy between $\what{\bB}_{1, \ell}$ and $\what{\bB}_{2, \ell}$. To stabilize numerical performance, we re-normalize $\hat{f}_\bU$ as 
\begin{equation}\label{eq:fk_eigenvector}
    f(\ell) = \frac{\hat{f}_\bU(\ell)}{1 + \sum_{\ell=1}^{L}\hat{f}_\bU(\ell)}, \quad \text{for } \ell = 1, 2, \ldots, L.
\end{equation}

Regarding eigenvalues, it is anticipated that $\hat{\lambda}_\nu$, estimated from the full dataset, will exhibit a steep decline at $\nu = d + 1$, transitioning from relatively large values observed at $\nu \leq d$. Similar to defining $f(\ell)$, we normalize the eigenvalues and define a function
\begin{equation}\label{eq:gk_eigenvalue}
    g(\ell) = \frac{\hat{\lambda}_\ell}{\sum_{\ell=1}^{L}\hat{\lambda}_\ell}, \quad\text{for } \ell = 1, 2, \ldots, L.
\end{equation}

\begin{figure}
    \centering
    \includegraphics[scale = 0.33]{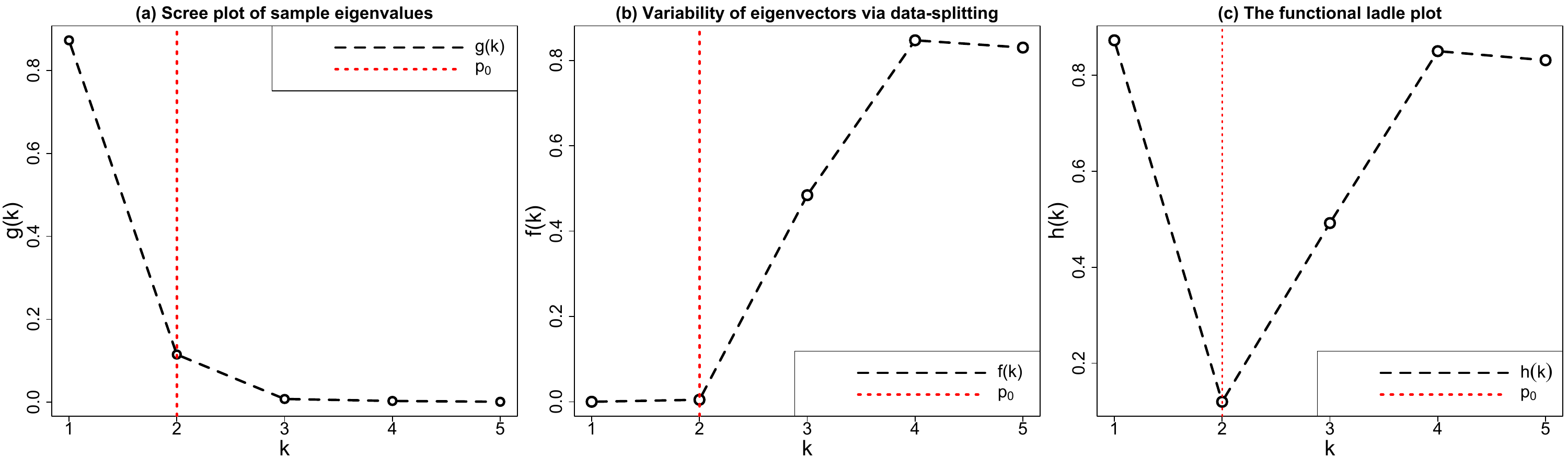}
    \caption{The leftmost plot presents $g(\ell)$, which illustrate the proportion of variation explained by each $\hat{\lambda}_\nu$. The middle panel shows $f(\ell)$, capturing the (normalized) variability between $\what{B}_{1, \ell}$ and $\what{B}_{2, \ell}$. The rightmost panel displays $h(\ell)$, which integrates information from both estimated eigenvalues and estimated eigenfunctions. In each panel, a vertical dotted line marks the true rank.}
    \label{fig:eigencomponents_plot}
\end{figure}

Combining the trends of $f(\ell)$ and $g(\ell)$, we construct a function characterized by a ``V'' shape, incorporating information from both the estimated eigenvalues and eigenfunctions. Specifically, the functional ladle estimator (FLE) is defined as 
\begin{equation*}\label{eq:hk_fle}
    h(\ell) = f(\ell) + g(\ell), \quad\text{for } \ell = 1, 2, \ldots, L,
\end{equation*}
with the components $f(\ell)$ and $g(\ell)$ specified in \eqref{eq:fk_eigenvector} and \eqref{eq:gk_eigenvalue}, respectively. In particular, $h(\ell)$ is expected to attain its minimum value around $\ell = d$. This intuition behind using $h(\ell)$ to estimate the rank is demonstrated in Figure \ref{fig:eigencomponents_plot}. 
From the leftmost panel of Figure \ref{fig:eigencomponents_plot}, which only accounts for the estimated eigenvalues, we may estimate the rank of $\what{G}$ as 2 or $3$. However, incorporating the variability of the estimated eigenfunctions, as shown in the middle panel of Figure \ref{fig:eigencomponents_plot}, suggests that the most plausible estimate is $\hat{d} = 2$. Algorithm \ref{algo:FLE_algo} provides the details for implementing our method to estimate the rank of functional data. 

\begin{algorithm}[!ht]
  \caption{Proposed order determination procedure}
  \vspace{1em}
  \begin{enumerate}
    \item Randomly split the data set $\D$ into two subsets $\D_1$ and $\D_2$.
    \item Obtain $\hat{\mu}_g, \what{G}_g, \{\hat{\lambda}_{g, \nu}, \hat{\phi}_{g, \nu}\}_{\nu = 1}^{L}$ based on the methods discussed in Section \ref{sec:FPCA} for each subset $\D_g, g = 1, 2$, where $L$ is a pre-specified number greater than $d$.
    \item Approximate $f_{\bU}(\ell)$ by $1 - \abs{\det(\delta \what{\bB}_{1,\ell}^{\top}\what{\bB}_{2,\ell})}$ and obtain $f(\ell)$ by equation \eqref{eq:fk_eigenvector} for all $\ell \leq L$.
    \item Obtain $\hat{\lambda}_{\nu}$ from $\D$ for all $\nu \leq L$, and compute $g(\ell)$ based on equation \eqref{eq:gk_eigenvalue}.
    \item Obtain the estimate of the rank by $\hat{d} = \underset{\ell \leq L}{\amin}~h(\ell) = \underset{\ell \leq L}{\amin}\brac{f(\ell) + g(\ell)}$.
\end{enumerate}
\label{algo:FLE_algo}
\end{algorithm}
\section{SIMULATION STUDIES}\label{sec:simulation}
In this section, we perform simulation studies to investigate the finite sample performance of the proposed method. 
We generate data $\{(Y_{ij}, T_{ij}) \mid i = 1, 2, \ldots, n, j = 1, 2,\ldots, N_i\}$ based on the Karhunen–Lo\`{e}ve expansion in \eqref{eq:KL_expansion_truncated} and model \eqref{eq:model}, where the true mean function is given by $\mu(t) = t + 10\exp\{-(t-1/2)^2\}$ for $t \in [0, 1]$. To showcase the performance under various sampling frequencies, we consider the number of observations per subject $N_i = m \in \{11, 26, 51\}$, where $m=11$ and $51$ are referred to as sparse and dense functional data, respectively. The intermediate case, $m = 26$, represents a transitional state between sparse and dense, hereby referred to as neither. Observation times $T_{ij}$ are uniformly generated over $[0, 1]$, referred to as irregular. In addition, for the dense case alone, we also examine a scenario where the data are collected at regularly spaced time points, referred to as regular.

Throughout this section, eigenfunctions are given by
\begin{gather*}
    \phi_1(t) = 1,~\phi_{2k}(t) = \sqrt{2}\sin(2k \pi t),~\phi_{2k + 1}(t) = \sqrt{2}\cos(2k \pi t) \quad k \in \N.
\end{gather*}
Regarding the true dimension $d$, eigenvalues $\lambda_\nu, \nu \in \N$, and the number of curves $n$, we consider the following scenarios. 
\begin{itemize}
    \item In the \textit{simple} setting, we consider $n  \in \{100, 200\}$, and $\lambda_\nu = (4-\nu)^2$ for $\nu = 1, 2, 3$, and $\lambda_\nu = 0$ for $\nu \geq 4$. Hence, $d = 3$. 
    \item In the \textit{complex} setting, we consider $n \in \{200, 300\}$, and $\lambda_\nu = (7-\nu)^2$ for $\nu = 1, 2, \ldots, 6$, and $\lambda_\nu = 0$ for all $\nu \geq 7$. Thus, $d = 6$. 
\end{itemize}
Additionally, we consider two designs for FPC scores: Gaussian, where scores $\xi_{i\nu} \sim N(0, \lambda_{\nu})$, and non-Gaussian, where scores $\xi_{i\nu} $ are generated from the centered exponential distribution with variance $\lambda_{\nu}$ for all $\nu \leq d$, introducing skewness into the functional data. The variance of measurement error, $\sigma_{\eps}^2$, is set to $\{0.1, 0.5, 1, 4\}$. Notably, the largest value of $\sigma_\eps^2$ exceeds the smallest nonzero eigenvalue in each scenario mentioned above, while the smallest $\sigma_\eps^2$ mimics the low-noise regime described in \cite{Hall_Vial_2006}.

For irregularly sampled data, we estimate model components using the methods described in Section \ref{sec:model_compoenent_estimation}. If observations are observed at a regularly dense grid, i.e., $T_{ij} = t_j$ for all $j=1, 2, \ldots, m$, we estimate the mean function using the sample mean $\hat{\mu}(t_j) = n^{-1}\sum_{i=1}^{n}Y_{ij}$. For the covariance function, we first estimate the raw covariance function by $\what{R}(t_j, t_k) = n^{-1}\sum_{i=1}^{n}\{Y_{ij} - \hat{\mu}(t_j)\}\{Y_{ik} - \hat{\mu}(t_{k})\}$. Thus, $\what{G}(t_{j}, t_{k}) = \what{R}(t_j, t_k) - \hat{\sigma}_\eps^2 \bI_{m}$, where $\hat{\sigma}_\eps^2$ can be estimated by a difference-based estimator \citep{rice_bandwidth_1984}, and $\bI_{m}$ denotes the $m \times m$ identity matrix.

Following \cite{Yao_Muller_Wang_2005_FPCA},
we select $h_{\mu}$ to estimate $\mu$ by generalized cross-validation (GCV). The optimal
$h_{\mu}$ minimizes the GCV error. 
Similarly, the optimal bandwidth $h_G$ for estimating $G$ is also determined by GCV, except in the scenario where $m = 51$ observations are irregularly spaced. In this particular scenario, we set $h_G = n^{-1/5}/6$ (approximately 0.06 when $n=200$ and 0.05 when $n=300$) to reduce the computational cost, as the $h_G$'s selected by GCV range between $0.04$ and $0.07$. 

Each simulation setting is repeated 500 times, and we report the percentage of times the true rank is identified. We compare our proposed method (FLE) with several commonly used approaches for selecting the rank of functional data, particularly focusing on IC-based methods. We include pseudo-AIC from \cite{Yao_Muller_Wang_2005_FPCA} (denoted as $\text{AIC}_{\text{Yao}}$) and pseudo-BIC implemented in R package \textit{PACE} (denoted as $\text{BIC}_{\text{PACE}}$). Besides that, we consider the modified AIC and BIC proposed by \cite{li_selecting_2013}, denoted by $\text{AIC}_{\text{Li}}$ and $\text{BIC}_{\text{Li}}$, respectively. Simulation results for Gaussian processes with $d =3$ are summarized in Tables \ref{tab:Gau_Irregular_nonDense_simple} and \ref{tab:Gau_Dense_simple}. 
\begin{table}
\centering
\caption{Comparison of order determination for Gaussian random processes under irregular design with $d = 3$ and $m=11, 26$: entries in Columns 4–11 are the percentage of accurate order determination across 500 iterations.}
\begin{tabular}[t]{c @{\hspace{1em}} c @{\hspace{1em}} c c c c c @{\hspace{1em}} c c c c}
\hline
\hline\\[-1.5ex]
{Number of} & \multirow{2}[2]{*}{Methods}& $n$ &\multicolumn{4}{c}{$100$} & \multicolumn{4}{c}{$200$}\\[0.5ex]
\cline{3-11}\\[-1.5ex]
measurements& &$\sigma_{\eps}^2$ & $0.1$ & $0.5$ & $1$ & $4$ & $0.1$ & $0.5$ & $1$ & $4$\\[0.5ex]
\hline
\multirow{5}{*}{$m$ = 11} & $\textbf{FLE}$& & 0.886 & 0.888 & 0.904 & 0.858 & 0.922 & 0.922 & 0.912 & 0.898 \\
&$\text{AIC}_{\text{Li}}$& & 0.796 & 0.898 & 0.982 & 0.828 & 0.752 & 0.938 & 0.994 & 0.968 \\
&$\text{BIC}_{\text{Li}}$& & 0.000 & 0.000 & 0.000 & 0.000 & 0.000 & 0.000 & 0.000 & 0.000 \\
&$\text{AIC}_{\text{Yao}}$& &0.112 & 0.066 & 0.068 & 0.100 & 0.006 & 0.008 & 0.002 & 0.006 \\
&$\text{BIC}_{\text{PACE}}$& & 0.310 & 0.232 & 0.244 & 0.318 & 0.104 & 0.042 & 0.034 & 0.064 \\
\hline
\multirow{5}{*}{$m$ = 26} & $\textbf{FLE}$& & 0.962 & 0.968 & 0.954 & 0.960 & 0.962 & 0.980 & 0.972 & 0.962 \\
&$\text{AIC}_{\text{Li}}$& & 0.266 & 0.722 & 0.914 & 0.998 & 0.358 & 0.930 & 0.986 & 1.000 \\
&$\text{BIC}_{\text{Li}}$& & 0.000 & 0.000 & 0.000 & 0.000 & 0.000 & 0.000 & 0.000 & 0.000 \\
&$\text{AIC}_{\text{Yao}}$& & 0.000 & 0.000 & 0.000 & 0.000 & 0.000 & 0.000 & 0.000 & 0.000  \\
&$\text{BIC}_{\text{PACE}}$& &0.004 & 0.000 & 0.002 & 0.002 & 0.002 & 0.000 & 0.000 & 0.000 \\
\hline
\end{tabular}
\label{tab:Gau_Irregular_nonDense_simple}
\end{table}
\begin{table}
\centering
\caption{Comparison of order determination for Gaussian random processes with $d = 3$ and $m=51$: entries in Columns 4–11 are the percentage of accurate order determination across 500 iterations.}
\begin{tabular}[t]{c @{\hspace{1em}} c @{\hspace{1em}} c c c c c @{\hspace{1em}} c c c c}
\hline
\hline\\[-1.5ex]
{Number of} & \multirow{2}[2]{*}{Methods}& $n$ &\multicolumn{4}{c}{$100$} & \multicolumn{4}{c}{$200$}\\[0.5ex]
\cline{3-11}\\[-1.5ex]
measurements& &$\sigma_{\eps}^2$ & $0.1$ & $0.5$ & $1$ & $4$ & $0.1$ & $0.5$ & $1$ & $4$\\[0.5ex]
\hline
\multirow{5}{*}{$m$ = 51, irregular} & $\textbf{FLE}$& & 0.966 & 0.956 & 0.976 & 0.980 & 0.966 & 0.980 & 0.986 & 0.976 \\
&$\text{AIC}_{\text{Li}}$& & 0.192 & 0.744 & 0.926 & 1.000 & 0.250 & 0.922 & 0.994 & 1.000 \\
&$\text{BIC}_{\text{Li}}$& & 0.000 & 0.000 & 0.000 & 0.000 & 0.000 & 0.000 & 0.000 & 0.000 \\
&$\text{AIC}_{\text{Yao}}$& &0.000 & 0.000 & 0.000 & 0.000 & 0.000 & 0.000 & 0.000 & 0.000 \\
&$\text{BIC}_{\text{PACE}}$& &0.000 & 0.000 & 0.000 & 0.000 & 0.000 & 0.000 & 0.000 & 0.000 \\
\hline
\multirow{5}{*}{$m$ = 51, regular} & $\textbf{FLE}$& & 1.000 & 1.000 & 1.000 & 1.000 & 1.000 & 1.000 & 1.000 & 1.000 \\
&$\text{AIC}_{\text{Li}}$& & 0.364 & 0.014 & 0.000 & 0.000 & 0.746 & 0.206 & 0.048 & 0.000 \\
&$\text{BIC}_{\text{Li}}$& & 1.000 & 1.000 & 1.000 & 1.000 & 1.000 & 1.000 & 1.000 & 1.000 \\
&$\text{AIC}_{\text{Yao}}$& & 0.038 & 0.000 & 0.000 & 0.000 & 0.062 & 0.000 & 0.000 & 0.000  \\
&$\text{BIC}_{\text{PACE}}$& &0.040 & 0.000 & 0.000 & 0.000 & 0.062 & 0.000 & 0.000 & 0.000 \\
\hline
\end{tabular}
\label{tab:Gau_Dense_simple}
\end{table}
Additional simulation results, including those for non-Gaussian processes and $d=6$, are provided in \ref{appendix:extra_simulation_results}.

The simulation results demonstrate that FLE is highly competitive across all simulation settings. In particular, when data are regularly and densely observed from a Gaussian process, FLE consistently identifies the true rank. Even in sparse settings with $m=11$, this method remains robust, achieving an accuracy of at least $86\%$ across all scenarios. Moreover, the performance improves as $m$ or $n$ increases, which implies that our method performs better when more aggregated observations are available. 

In contrast, IC-based methods exhibit high instability. For example, the accuracy of $\text{AIC}_{\text{Li}}$ ranges from $20\%$ to $100\%$ depending on the value of $\sigma_\eps^2$ for dense functional data. This instability is evident in every setting that we have examined. Moreover, the performance of IC-based methods that are developed based on likelihood functions is highly sensitive to the values of $n, m$ and $\sigma_{\eps}^2$. Our numerical analysis reveals that these methods perform well only for specific parameter combinations. Furthermore, although $\text{BIC}_{\text{Li}}$ does not depend on the likelihood function, it requires accurate estimation of eigenvalues, particularly for those with indices close to the true rank $d$. Consequently, its performance deteriorates under the setting of irregular and sparse observations or a high noise variance $\sigma_{\eps}^2$, since accurate estimation of eigenvalues with indices close to $d$ becomes quite challenging.

\section{REAL DATA EXAMPLES}\label{sec:real_data}
In this section, we apply our proposed method to two real-world examples. In reality, we can never know the true order of functional data, thus we cannot evaluate the accuracy of our method in estimating the dimension of functional data directly. Instead, we illustrate its effectiveness indirectly. In the literature on functional linear regression and functional data classification, FPC-based methods have received extensive attention; see \cite{Yao_Muller_Wang_2005_regression}, \cite{hall_methodology_2007}, \cite{delaigle2012achieving}, \cite{dai2017optimal} and references therein.
More specifically, FPC scores obtained from FPCA are used as covariates in these regression and classification methods. Selecting the number of FPC scores is a critical problem when applying these methods. We demonstrate the effectiveness of our method by showcasing its performance in FPC-based methods for functional linear regression and functional data classification.

\subsection{Capital Bikeshare Data}
We analyze a dataset from the Capital Bikeshare System (CBS) in Washington, D.C., to investigate the relationship between the hourly rental profile and the total rental time (TRT) in hours on the same day. This relationship serves as a potential commercial indicator for assessing whether bicycle demand exceeds supply in a given region. The CBS dataset records the date and time of each rental, offering valuable insights into public transportation usage and environmental factors.

For this study, we consider rental transactions for the year 2017, restricting our analysis to users with memberships. Given that rental patterns differ significantly between weekdays and weekends, we focus solely on weekends, including holidays, resulting in a total of 116 days of data. Additionally, for each day, we remove records where rental durations exceed five hours, as these extremely long rental times are likely due to users forgetting to return bikes.

Our objective is to formally assess how the hourly rental profile, denoted as $X(\cdot)$, affect the TRT, denoted as $Y$, for each day. Figure \ref{fig:the_number_of_rentals_over_time} displays the hourly rental profiles across all 116 days, and Figure \ref{fig:total_rental_time} presents the histogram of rental durations (both available in \ref{appendix:extra_real_data_analysis}). We consider a scalar-on-function regression model to quantify the relationship between $X$ and $Y$, which is defined as follows:
\begin{equation}\label{eq:sofr_model_real_data}
    Y_i = \alpha + \int \beta(t)X_i(t)dt + e_i,\quad i \in \{1, 2, \ldots, 249\},
\end{equation}
where the coefficient function $\beta(\cdot)$ quantifies the impact of hourly rental numbers on TRT. 

We employ the FPC-based method developed by \cite{hall_methodology_2007} to fit model \eqref{eq:sofr_model_real_data}. 
To compare the performance of the proposed method with other IC-based methods, we randomly split the dataset into $90\%$ training data and $10\%$ test data. Since $X$ is observed on a common grid, we estimate the mean function and covariance function using sample averages and the empirical covariance matrix on the training data. Let $\hat{d}$ denote the number of FPCs selected by a selection algorithm. Applying FPCA to $X$, model \eqref{eq:sofr_model_real_data} reduces to a linear regression model:
$$Y_i =  \beta_0 + \sum_{\nu=1}^{\hat{d}}\beta_{\nu}\xi_{i\nu} + e_i,$$
where $\beta_0 = \alpha + \int_{t}\beta(t)\mu(t)dt$ and $\beta_{\nu} = \int_{t}\beta(t)\phi_{\nu}(t)dt$ for $\nu \geq 1$. After obtaining the estimate of $\boldsymbol{\beta} = (\beta_0, \beta_1, \beta_2, \ldots, \beta_{\hat{d}})^{\top}$ by ordinary least squares, the estimate of $\beta$ in model \eqref{eq:sofr_model_real_data} is given by $\hat{\beta}(t) = \sum_{\nu=1}^{\hat{d}}\hat{\beta}_{\nu}\hphi_{\nu}(t)$, where $\phi_{\nu}$ denote the estimated eigenfunctions of $X$. 

Using the estimated slope function $\hat{\beta}(t)$, we predict TRT for days in the test set, and compute the prediction error, defined as
$\sum_{i \in \I}(y_i - \hat{y}_i)^2/\abs{\I}$, where $\I$ is the index set for test days and $\abs{\I}$ denotes the total number of test days. Table \ref{tab:bikeShare_data} summarizes the estimated rank and the corresponding prediction errors for the proposed method and IC-based methods across 500 independent splits. 
\begin{table}
    \centering
    \caption{The estimated rank of the bike sharing data with averaged prediction errors under 500 runs. The estimated rank is the mode of estimated ranks.}
    \vspace{1.1em}
    \begin{tabular}{c c c c c c}
        \hline
        \hline
        method & FLE & AIC$_{\text{Yao}}$ & BIC$_{\text{PACE}}$ & AIC$_{\text{m}}$ & BIC$_{\text{m}}$ \\
        \hline
        $\hat{d}$ & 4 & 22 & 22 & 22 & 21\\
        \hline
        prediction error & 7584.569 & 7822.712 & 7822.712 & 7822.712 & 7903.395\\
        \hline
    \end{tabular}
    \label{tab:bikeShare_data}
\end{table}
The results demonstrate the superiority of our method in determining the order of functional data, as indicated by lower prediction errors. The estimated slope function based on the first four eigenfunctions, the first three estimated eigenfunctions, and additional estimated eigenfunctions, are displayed in Figures \ref{fig:bike_slope_function} - \ref{fig:bike_eigenfunctions_extra} in \ref{appendix:extra_real_data_analysis}.

\subsection{Beijing Air Pollutants Data}
Air pollution has become a major environmental concern in many cities across China due to rapid industrialization and urbanization. Fine particulate matter (PM) with an aerodynamic diameter less than $\SI{2.5}{\micro\meter}$, often referred to as PM$_{2.5}$, is one of the main pollutants in Beijing \citep{Zhang2023Envir}. Exposure to PM$_{2.5}$ has been associated with cardiovascular and respiratory diseases and even lung cancer \citep{PopeJAMA2002, hoek_long-term_2013, Lelieveld2015Nature}.

As noted in \cite{LiangJRSSA2015} and \cite{Zhang2023Envir}, PM$_{2.5}$ levels are highly influenced by meteorological conditions, particularly wind and humidity conditions. In this example, we focus on the effect of dew point temperature (DEW), which serves as a proxy for both relative humidity and air temperature \citep{DEWP_1996}. Moreover, DEW has been recognized as an important predictor for the levels of PM$_{2.5}$. For example, the first branching rule of a tree model proposed in \cite{Zhang_Guo_Dong_He_Xu_Chen_2017} is decided by DEW, and its selection frequency is $100\%$ in the selection frequency chart of \cite{LiangJRSSA2015}.

To demonstrate the effectiveness of the proposed method, we use DEW to predict the daily PM$_{2.5}$ level, denoted as $Y$, whose label is based on the daily average readings of PM$_{2.5}$. Specifically, daily average readings are divided into four categories: $Y = 0$ if PM$_{2.5} \leq \SI{35}{\micro\gram\meter^{-3}},~ Y = 1$ if $\SI{35}{\micro\gram\meter^{-3}} < $ PM$_{2.5} \leq \SI{75}{\micro\gram\meter^{-3}},~Y = 2$ if $\SI{75}{\micro\gram\meter^{-3}} < $ PM$_{2.5} \leq \SI{150}{\micro\gram\meter^{-3}},~ Y = 3$, if $\SI{150}{\micro\gram\meter^{-3}} < $ PM$_{2.5}$. This partition adapts the current national ambient air quality standard in China (\url{https://www.mee.gov.cn/ywgz/fgbz/bz/bzwb/dqhjbh/dqhjzlbz/201203/W020120410330232398521.pdf}, in Chinese), which is also adopted in \cite{Zhang_Guo_Dong_He_Xu_Chen_2017}.

To predict $Y$ with DEW, we consider the following generalized scalar-on-function regression:
\begin{equation*}
    \log \frac{\Pr(Y=k \mid X)}{\Pr(Y = 3 \mid X)} = \alpha_{k} + \int_{t}\beta_{k}(t)X(t)dt + e \quad\text{for } k = 0, 1, 2, 
\end{equation*}
where $X(t)$ denotes the hourly DEW readings.
Such models with a general link function were studied in \cite{muller2005generalized}. 
We analyze hourly air pollution data collected from Huairou, an urban district in the northern part of Beijing, spanning March 2013 to February 2017. To improve prediction accuracy, we stratify the data based on human activity patterns and seasonal effects. First, we separate workdays from weekends and holidays, ensuring that each subset exhibits homogeneous human activity patterns. Given the strong seasonal variability of air pollution levels in Beijing \citep{Zhang_Guo_Dong_He_Xu_Chen_2017}, we partition 12 months into four seasons. In this study, we consider winter months (December–February of the following year) as winter heating is one of the most important factors directly impacting PM$_{2.5}$ levels \citep{Zhang2023Envir}. This partitioning strategy is widely used in air pollution studies \citep{LiangJRSSA2015}. Finally, to evaluate the prediction accuracy, we randomly select $10\%$ of the data as a test set while using the remaining $90\%$ for training.

In the training set, we estimate the mean function and covariance function of $X$ using the same procedures as in the Capital Bikeshare study since DEW is recorded on a common grid. The dimension $\hat{d}$ is determined by the proposed method and other IC-based methods. Using the estimated slope function and the FPC scores, we predict $Y$ in the test set and evaluate classification accuracy on the test set to compare the effectiveness of different methods in estimating $d$.

Table \ref{tab:beijing_data_workday} summarizes the estimated rank and the classification accuracy for winter workdays over 500 independent runs.
\begin{table}
    \centering
    \caption{The estimated rank of the Beijing air pollutants data and the classification accuracy for different methods in winter workdays under 500 runs. The estimated rank is the mode of estimated ranks.}
    \vspace{0.5em}
    \begin{tabular}{c c c c c c}
        \hline
        \hline
        method & FLE & AIC$_{\text{Yao}}$ & BIC$_{\text{PACE}}$ & AIC$_{\text{Li}}$ & BIC$_{\text{Li}}$\\
        \hline
        $\hat{d}$ & 4 & 11 & 11 & 4 & 13\\
        \hline
        accuracy & 0.542 & 0.522 & 0.523 & 0.542 & 0.509\\
        \hline
        \hline
    \end{tabular}
    \label{tab:beijing_data_workday}
\end{table}
The results indicate that our method achieves the highest classification accuracy, demonstrating its superior predictive performance in classifying functional data, and thus its great capability to estimate the order of functional data. Furthermore, our analysis suggests that DEW is a strong predictor of PM$_{2.5}$ pollution levels, consistent with prior studies \citep{Zhang_Guo_Dong_He_Xu_Chen_2017}.
\section{CONCLUSION}\label{sec:conc}
In this paper, we develop a novel procedure for determining the order of functional data when the corresponding covariance operator has finite rank. Our method does not rely on the Gaussian assumption for estimating FPC scores or the low-noise regime, making it more flexible and widely applicable. Numerical studies demonstrate the strong performance of the proposed method across various settings, whereas IC-based methods often exhibit sensitivity to specific parameter choices.

Despite these promising results, several theoretical aspects remain to be explored. In particular, analyzing discrete contaminated observations from infinite-dimensional processes presents significant challenges. A key direction for future research is to rigorously characterize the asymptotic behavior of the estimated eigenfunctions in the null space of the covariance operator. In future work, we aim to investigate these theoretical properties in greater depth to further enhance the robustness and applicability of our approach.

\addcontentsline{toc}{section}{Appendices}
\setcounter{equation}{0}
\setcounter{figure}{0}
\setcounter{table}{0}
\renewcommand{\theequation}{\Alph{section}.\arabic{equation}}
\renewcommand{\thefigure}{\Alph{section}.\arabic{figure}}
\renewcommand{\thetable}{\Alph{section}.\arabic{table}}
\renewcommand{\baselinestretch}{1.9}

\appendix
\renewcommand{\appendixname}{Appendix}
\renewcommand{\thesection}{Appendix \Alph{section}} 
\section{Estimated eigenfunctions}\label{appendix:FLE_sec}
\begin{figure}[!hb]
    \centering
    \includegraphics[scale = 0.65]{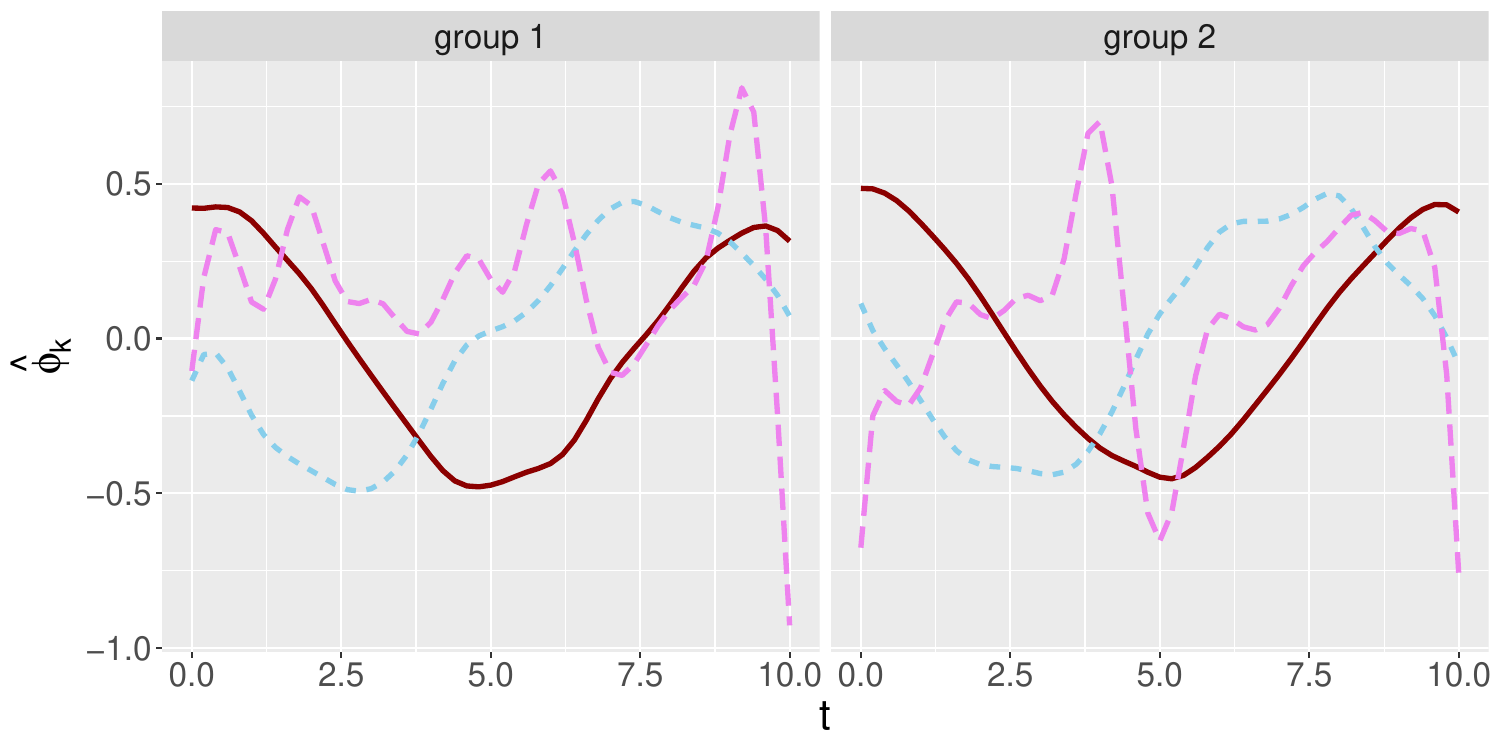}
    \caption{The left panel displays the first three estimated eigenfunctions derived from $\D_1$, while the right panel presents those from $\D_2$. Here the solid line represents $\hphi_1$, the dash line is $\hphi_2$, and the long dash line is $\hphi_3$, for both panels.} 
    \label{fig:first_three_eigenfuncs}
\end{figure}

We generate the data $\D = (Y_{ij}, T_{ij})$ based on the model \eqref{eq:model} and the KL expansion in equation \eqref{eq:KL_expansion_truncated}, where $\mu(t) = t + 10\exp\{-(t-5)^2\}$ for $t \in [0, 10]$ and $\sigma_\eps^2 = 0.01$. The eigenfunctions are given by
\begin{gather*}
    \phi_1(t) = 5^{-1/2}\cos(\pi t/5),~\phi_{2}(t) = -5^{-1/2}\sin(\pi t/5)\quad k \in \N,
\end{gather*}
and $\lambda_1 = 25, \lambda_2 = 4$ and $\lambda_\nu = 0$ for $\nu \geq 3$. Thus, the dimension of $X$ is $2$. Figure \ref{fig:first_three_eigenfuncs} illustrates the variability of estimated eigenfunctions based on one simulation run. From the figure, we can see that the first two estimated eigenfunctions from each subset appear similar in shape, whereas the third estimated eigenfunction deviates notably.
\section{Additional simulation results}\label{appendix:extra_simulation_results}
This section provides extra simulation studies results. Tables \ref{tab:Gau_Irregular_nonDense_complex} and \ref{tab:Gau_Dense_complex} show the results when $d = 6$ and the underlying processes are Gaussian. Table \ref{tab:nonGau_Irregular_nonDense_simple} - Table \ref{tab:nonGau_Dense_complex} illustrate the results for non-Gaussian processes, where $d = 3$ for first two tables and $d = 6$ for the last two tables.
\begin{table}[!ht]
\centering
\caption{Comparison of order determination for Gaussian random processes under irregular design with $d = 6$ and $m=11, 26$: entries in Columns 4–11 are the percentage of accurate order determination across 500 iterations.}
\begin{tabular}[t]{c @{\hspace{1em}} c @{\hspace{1em}} c c c c c @{\hspace{1em}} c c c c}
\hline
\hline
{Number of} & \multirow{2}[2]{*}{Methods}& $n$ &\multicolumn{4}{c}{$200$} & \multicolumn{4}{c}{$300$}\\[0.8ex]
\cline{3-11}
measurements& &$\sigma_{\eps}^2$ & $0.1$ & $0.5$ & $1$ & $4$ & $0.1$ & $0.5$ & $1$ & $4$\\[0.5ex]
\hline
\multirow{5}{*}{$m$ = 11} & $\textbf{FLE}$& & 0.112 & 0.120 & 0.124 & 0.096 & 0.158 & 0.148 & 0.164 & 0.152 \\
&$\text{AIC}_{\text{Li}}$& & 0.126 & 0.102 & 0.084 & 0.016 & 0.256 & 0.192 & 0.176 & 0.050 \\
&$\text{BIC}_{\text{Li}}$& & 0.000 & 0.000 & 0.000 & 0.000 & 0.000 & 0.000 & 0.000 & 0.000 \\
&$\text{AIC}_{\text{Yao}}$& &0.198 & 0.218 & 0.218 & 0.230 & 0.064 & 0.104 & 0.094 & 0.158 \\
&$\text{BIC}_{\text{PACE}}$& &0.550 & 0.496 & 0.522 & 0.468 & 0.328 & 0.320 & 0.274 & 0.320 \\
\hline
\multirow{5}{*}{$m$ = 26} & $\textbf{FLE}$& & 0.200 & 0.204 & 0.170 & 0.162 & 0.220 & 0.242 & 0.208 & 0.172 \\
&$\text{AIC}_{\text{Li}}$& & 0.258 & 0.370 & 0.468 & 0.734 & 0.206 & 0.348 & 0.466 & 0.780 \\
&$\text{BIC}_{\text{Li}}$& & 0.000 & 0.000 & 0.000 & 0.000 & 0.000 & 0.000 & 0.000 & 0.000 \\
&$\text{AIC}_{\text{Yao}}$& & 0.000 & 0.008 & 0.006 & 0.002 & 0.000 & 0.000 & 0.000 & 0.000  \\
&$\text{BIC}_{\text{PACE}}$& &0.012 & 0.026 & 0.014 & 0.016 & 0.014 & 0.020 & 0.012 & 0.000 \\
\hline
\end{tabular}
\label{tab:Gau_Irregular_nonDense_complex}
\end{table}
\begin{table}[!ht]
\centering
\caption{Comparison of order determination for Gaussian random processes with $d = 6$ and $m=51$: entries in Columns 4–11 are the percentage of accurate order determination across 500 iterations.}
\begin{tabular}[t]{c @{\hspace{1em}} c @{\hspace{1em}} c c c c c @{\hspace{1em}} c c c c}
\hline
\hline
{Number of} & \multirow{2}[2]{*}{Methods}& $n$ &\multicolumn{4}{c}{$200$} & \multicolumn{4}{c}{$300$}\\[0.8ex]
\cline{3-11}
measurements& &$\sigma_{\eps}^2$ & $0.1$ & $0.5$ & $1$ & $4$ & $0.1$ & $0.5$ & $1$ & $4$\\[0.5ex]
\hline
\multirow{5}{*}{$m$ = 51, irregular} & $\textbf{FLE}$& & 0.632 & 0.622 & 0.626 & 0.590 & 0.744 & 0.762 & 0.742 & 0.734 \\
&$\text{AIC}_{\text{Li}}$& & 0.572 & 0.670 & 0.828 & 0.956 & 0.390 & 0.676 & 0.798 & 0.982 \\
&$\text{BIC}_{\text{Li}}$& & 0.000 & 0.000 & 0.000 & 0.000 & 0.000 & 0.000 & 0.000 & 0.000 \\
&$\text{AIC}_{\text{Yao}}$& &0.002 & 0.002 & 0.000 & 0.000 & 0.000 & 0.000 & 0.000 & 0.000 \\
&$\text{BIC}_{\text{PACE}}$& &0.152 & 0.106 & 0.062 & 0.028 & 0.032 & 0.010 & 0.010 & 0.000 \\
\hline
\multirow{5}{*}{$m$ = 51, regular} & $\textbf{FLE}$& & 1.000 & 1.000 & 1.000 & 1.000 & 1.000 & 1.000 & 1.000 & 1.000 \\
&$\text{AIC}_{\text{Li}}$& & 1.000 & 0.960 & 0.624 & 0.082 & 1.000 & 1.000 & 0.986 & 0.462 \\
&$\text{BIC}_{\text{Li}}$& & 0.000 & 0.216 & 0.344 & 0.366 & 0.000 & 0.410 & 0.590 & 0.680 \\
&$\text{AIC}_{\text{Yao}}$& & 0.606 & 0.074 & 0.004 & 0.000 & 0.566 & 0.100 & 0.012 & 0.000  \\
&$\text{BIC}_{\text{PACE}}$& &0.608 & 0.080 & 0.004 & 0.000 & 0.568 & 0.100 & 0.012 & 0.000 \\
\hline
\end{tabular}
\label{tab:Gau_Dense_complex}
\end{table}
\begin{table}[!hb]
\centering
\caption{Comparison of order determination for non-Gaussian random processes under irregular design with $d = 3$ and $m=11, 26$: entries in Columns 4–11 are the percentage of accurate order determination across 500 iterations.}
\begin{tabular}[t]{c @{\hspace{1em}} c @{\hspace{1em}} c c c c c @{\hspace{1em}} c c c c}
\hline
\hline
{Number of} & \multirow{2}[2]{*}{Methods}& $n$ &\multicolumn{4}{c}{$100$} & \multicolumn{4}{c}{$200$}\\[0.8ex]
\cline{3-11}
measurements& &$\sigma_{\eps}^2$ & $0.1$ & $0.5$ & $1$ & $4$ & $0.1$ & $0.5$ & $1$ & $4$\\[0.5ex]
\hline
\multirow{5}{*}{$m$ = 11} & $\textbf{FLE}$& & 0.848 & 0.814 & 0.818 & 0.770 & 0.888 & 0.890 & 0.912 & 0.882 \\
&$\text{AIC}_{\text{Li}}$& & 0.822 & 0.868 & 0.872 & 0.596 & 0.772 & 0.888 & 0.952 & 0.818 \\
&$\text{BIC}_{\text{Li}}$& & 0.000 & 0.000 & 0.000 & 0.000 & 0.000 & 0.000 & 0.000 & 0.000 \\
&$\text{AIC}_{\text{Yao}}$& &0.346 & 0.362 & 0.296 & 0.372 & 0.132 & 0.120 & 0.112 & 0.118 \\
&$\text{BIC}_{\text{PACE}}$& &0.548 & 0.504 & 0.488 & 0.594 & 0.290 & 0.260 & 0.260 & 0.336 \\
\hline
\multirow{5}{*}{$m$ = 26} & $\textbf{FLE}$& & 0.946 & 0.934 & 0.950 & 0.940 & 0.972 & 0.956 & 0.964 & 0.946 \\
&$\text{AIC}_{\text{Li}}$& & 0.278 & 0.604 & 0.756 & 0.976 & 0.218 & 0.642 & 0.832 & 0.984 \\
&$\text{BIC}_{\text{Li}}$& & 0.000 & 0.000 & 0.000 & 0.000 & 0.000 & 0.000 & 0.000 & 0.000 \\
&$\text{AIC}_{\text{Yao}}$& & 0.014 & 0.014 & 0.010 & 0.022 & 0.000 & 0.002 & 0.000 & 0.006  \\
&$\text{BIC}_{\text{PACE}}$& &0.056 & 0.042 & 0.034 & 0.038 & 0.008 & 0.006 & 0.002 & 0.012 \\
\hline
\end{tabular}
\label{tab:nonGau_Irregular_nonDense_simple}
\end{table}
\begin{table}[!ht]
\centering
\caption{Comparison of order determination for non-Gaussian random processes with $d = 3$ and $m=51$: entries in Columns 4–11 are the percentage of accurate order determination across 500 iterations.}
\begin{tabular}[t]{c @{\hspace{1em}} c @{\hspace{1em}} c c c c c @{\hspace{1em}} c c c c}
\hline
\hline
{Number of} & \multirow{2}[2]{*}{Methods}& $n$ &\multicolumn{4}{c}{$100$} & \multicolumn{4}{c}{$200$}\\[0.8ex]
\cline{3-11}
measurements& &$\sigma_{\eps}^2$ & $0.1$ & $0.5$ & $1$ & $4$ & $0.1$ & $0.5$ & $1$ & $4$\\[0.5ex]
\hline
\multirow{5}{*}{$m$ = 51, irregular} & $\textbf{FLE}$& & 0.978 & 0.974 & 0.964 & 0.962 & 0.984 & 0.978 & 0.972 & 0.978 \\
&$\text{AIC}_{\text{Li}}$& & 0.094 & 0.448 & 0.684 & 0.972 & 0.086 & 0.528 & 0.774 & 0.994 \\
&$\text{BIC}_{\text{Li}}$& & 0.000 & 0.000 & 0.000 & 0.000 & 0.000 & 0.000 & 0.000 & 0.000 \\
&$\text{AIC}_{\text{Yao}}$& &0.002 & 0.002 & 0.000 & 0.000 & 0.000 & 0.000 & 0.000 & 0.000 \\
&$\text{BIC}_{\text{PACE}}$& &0.002 & 0.002 & 0.000 & 0.000 & 0.000 & 0.000 & 0.000 & 0.000 \\
\hline
\multirow{5}{*}{$m$ = 51, regular} & $\textbf{FLE}$& & 1.000 & 1.000 & 1.000 & 1.000 & 1.000 & 1.000 & 1.000 & 1.000 \\
&$\text{AIC}_{\text{Li}}$& & 0.338 & 0.010 & 0.000 & 0.000 & 0.750 & 0.206 & 0.058 & 0.000 \\
&$\text{BIC}_{\text{Li}}$& & 1.000 & 1.000 & 1.000 & 1.000 & 1.000 & 1.000 & 1.000 & 1.000 \\
&$\text{AIC}_{\text{Yao}}$& & 0.040 & 0.000 & 0.000 & 0.000 & 0.078 & 0.000 & 0.000 & 0.000  \\
&$\text{BIC}_{\text{PACE}}$& &0.042 & 0.000 & 0.000 & 0.000 & 0.080 & 0.000 & 0.000 & 0.000  \\
\hline
\end{tabular}
\label{tab:nonGau_Dense_simple}
\end{table}
\begin{table}[!hb]
\centering
\caption{Comparison of order determination for non-Gaussian random processes under irregular design with $d = 6$ and $m=11, 26$: entries in Columns 4–11 are the percentage of accurate order determination across 500 iterations.}
\begin{tabular}[t]{c @{\hspace{1em}} c @{\hspace{1em}} c c c c c @{\hspace{1em}} c c c c}
\hline
\hline
{Number of} & \multirow{2}[2]{*}{Methods}& $n$ &\multicolumn{4}{c}{$200$} & \multicolumn{4}{c}{$300$}\\[0.8ex]
\cline{3-11}
measurements& &$\sigma_{\eps}^2$ & $0.1$ & $0.5$ & $1$ & $4$ & $0.1$ & $0.5$ & $1$ & $4$\\[0.5ex]
\hline
\multirow{5}{*}{$m$ = 11} & $\textbf{FLE}$& & 0.066 & 0.058 & 0.036 & 0.050 & 0.096 & 0.082 & 0.058 & 0.096 \\
&$\text{AIC}_{\text{Li}}$& & 0.046 & 0.022 & 0.024 & 0.010 & 0.060 & 0.036 & 0.040 & 0.010 \\
&$\text{BIC}_{\text{Li}}$& & 0.000 & 0.000 & 0.000 & 0.000 & 0.000 & 0.000 & 0.000 & 0.000 \\
&$\text{AIC}_{\text{Yao}}$& &0.550 & 0.568 & 0.552 & 0.550 & 0.458 & 0.498 & 0.502 & 0.470 \\
&$\text{BIC}_{\text{PACE}}$& &0.428 & 0.382 & 0.384 & 0.418 & 0.610 & 0.528 & 0.580 & 0.592 \\
\hline
\multirow{5}{*}{$m$ = 26} & $\textbf{FLE}$& & 0.174 & 0.162 & 0.142 & 0.152 & 0.220 & 0.214 & 0.192 & 0.196 \\
&$\text{AIC}_{\text{Li}}$& & 0.570 & 0.570 & 0.596 & 0.520 & 0.642 & 0.652 & 0.654 & 0.666 \\
&$\text{BIC}_{\text{Li}}$& & 0.000 & 0.000 & 0.000 & 0.000 & 0.000 & 0.000 & 0.000 & 0.000 \\
&$\text{AIC}_{\text{Yao}}$& & 0.046 & 0.044 & 0.048 & 0.026 & 0.004 & 0.006 & 0.000 & 0.006  \\
&$\text{BIC}_{\text{PACE}}$& & 0.284 & 0.230 & 0.244 & 0.222 & 0.122 & 0.102 & 0.072 & 0.064 \\
\hline
\end{tabular}
\label{tab:nonGau_Irregular_nonDense_complex}
\end{table}
\begin{table}[!ht]
\centering
\caption{Comparison of order determination for Gaussian random processes with $d = 6$ and $m=51$: entries in Columns 4–11 are the percentage of accurate order determination across 500 iterations.}
\begin{tabular}[t]{c @{\hspace{1em}} c @{\hspace{1em}} c c c c c @{\hspace{1em}} c c c c}
\hline
\hline
{Number of} & \multirow{2}[2]{*}{Methods}& $n$ &\multicolumn{4}{c}{$200$} & \multicolumn{4}{c}{$300$}\\[0.8ex]
\cline{3-11}
measurements& &$\sigma_{\eps}^2$ & $0.1$ & $0.5$ & $1$ & $4$ & $0.1$ & $0.5$ & $1$ & $4$\\[0.5ex]
\hline
\multirow{5}{*}{$m$ = 51, irregular} & $\textbf{FLE}$& & 0.426 & 0.398 & 0.440 & 0.406 & 0.532 & 0.474 & 0.514 & 0.516 \\
&$\text{AIC}_{\text{Li}}$& & 0.396 & 0.542 & 0.668 & 0.850 & 0.252 & 0.478 & 0.614 & 0.870 \\
&$\text{BIC}_{\text{Li}}$& & 0.000 & 0.000 & 0.000 & 0.000 & 0.000 & 0.000 & 0.000 & 0.000 \\
&$\text{AIC}_{\text{Yao}}$& &0.006 & 0.000 & 0.000 & 0.000 & 0.000 & 0.000 & 0.000 & 0.000  \\
&$\text{BIC}_{\text{PACE}}$& &0.100 & 0.052 & 0.064 & 0.034 & 0.014 & 0.004 & 0.006 & 0.000 \\
\hline
\multirow{5}{*}{$m$ = 51, regular} & $\textbf{FLE}$& & 1.000 & 1.000 & 1.000 & 1.000 & 1.000 & 1.000 & 1.000 & 1.000 \\
&$\text{AIC}_{\text{Li}}$& & 1.000 & 0.952 & 0.600 & 0.070 & 1.000 & 1.000 & 0.978 & 0.522 \\
&$\text{BIC}_{\text{Li}}$& & 0.000 & 0.214 & 0.324 & 0.378 & 0.000 & 0.400 & 0.570 & 0.706 \\
&$\text{AIC}_{\text{Yao}}$& & 0.500 & 0.088 & 0.004 & 0.000 & 0.540 & 0.094 & 0.016 & 0.000  \\
&$\text{BIC}_{\text{PACE}}$& & 0.504 & 0.094 & 0.004 & 0.000 & 0.544 & 0.094 & 0.016 & 0.000 \\
\hline
\end{tabular}
\label{tab:nonGau_Dense_complex}
\end{table}

\section{Additional real data analysis}\label{appendix:extra_real_data_analysis}
\begin{figure}
    \centering
    \caption{The number of rentals over time, $X_i(\cdot)$}
    \includegraphics[scale = 0.5]{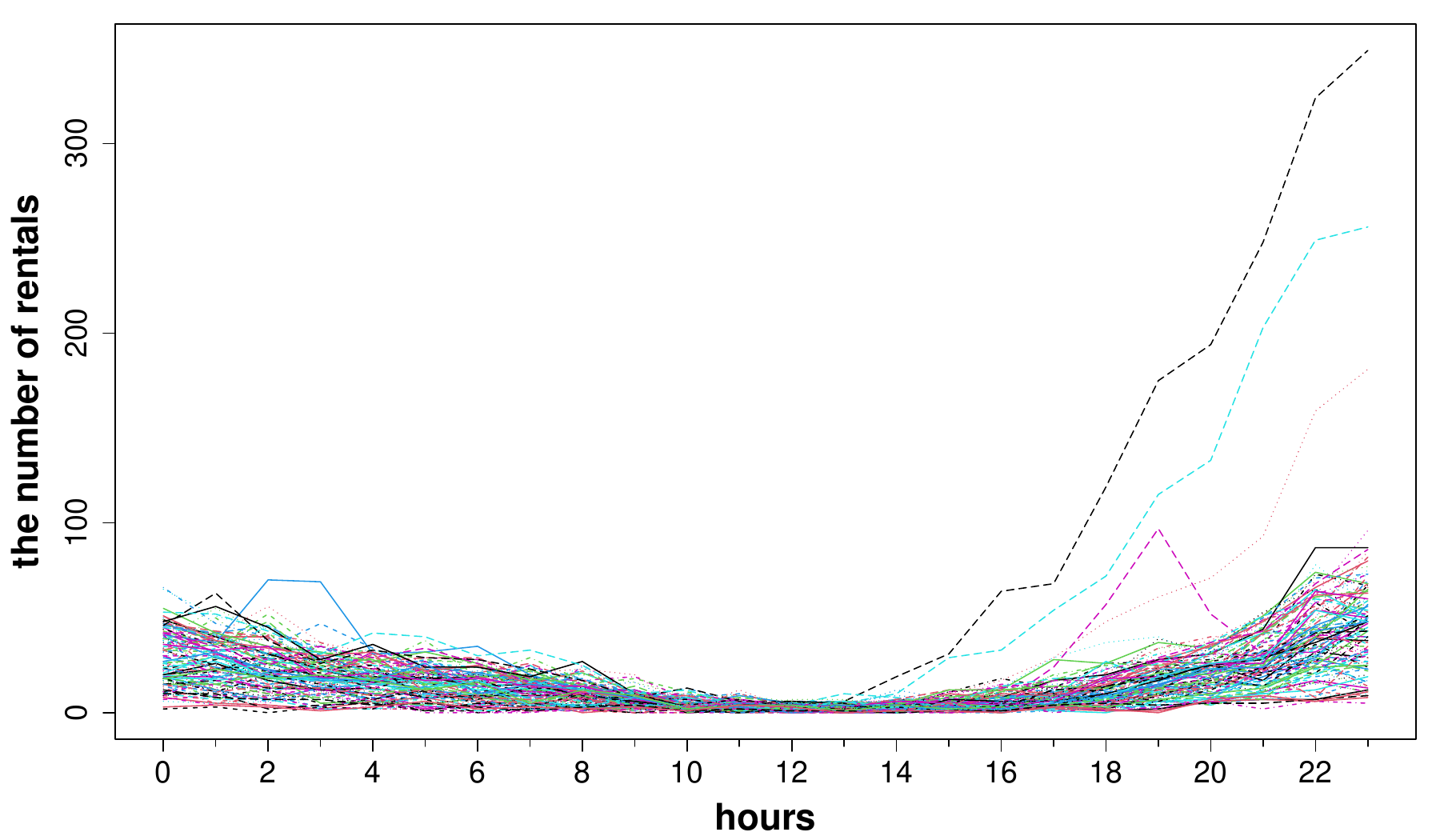}
    \label{fig:the_number_of_rentals_over_time}
\end{figure}

\begin{figure}[!ht]
    \centering
    \includegraphics[scale = 0.5]{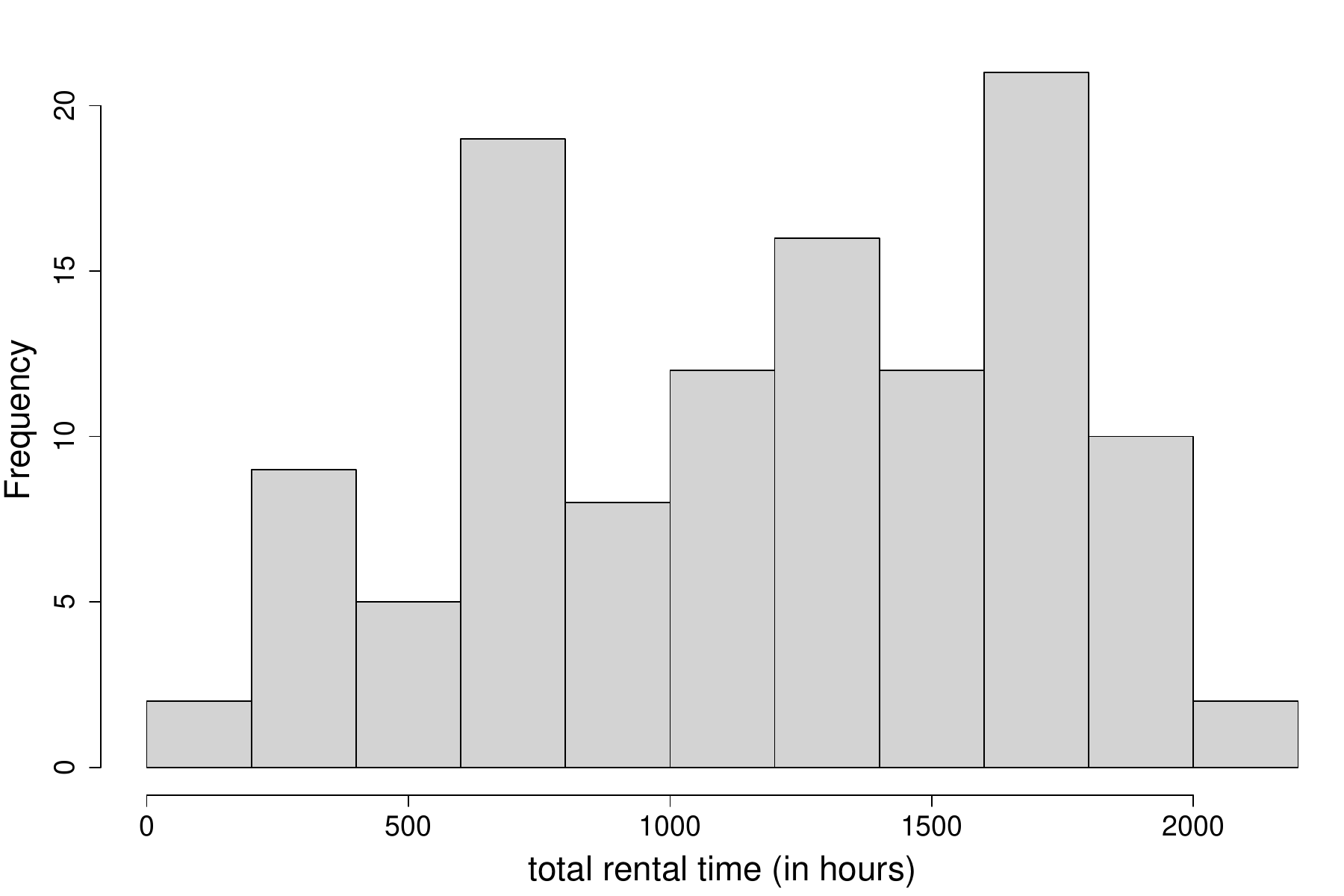}
    \caption{Total rental time in hours}
    \label{fig:total_rental_time}
\end{figure}

\begin{figure}
    \centering
    \includegraphics[scale = 0.5]{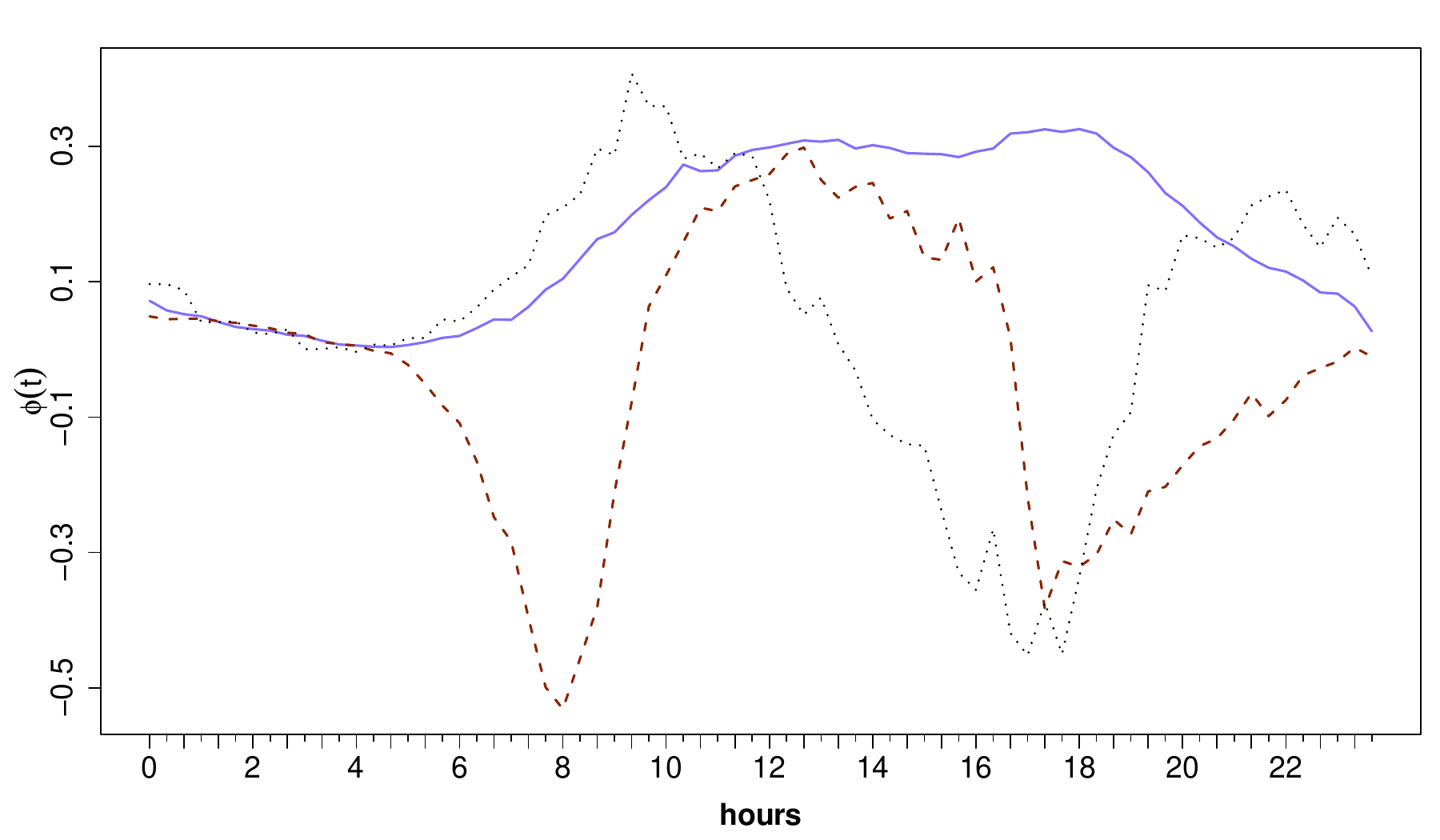}
    \caption{The plot shows the top three eigenfunctions for the bike-sharing data, where the first eigenfunction is presented in solid line, the second eigenfunction is in dash line, and the third eigenfunction is the dotted line.}
    \label{fig:bike_eigenfunctions}
\end{figure}

\begin{figure}
    \centering
    \includegraphics[width=\textwidth]{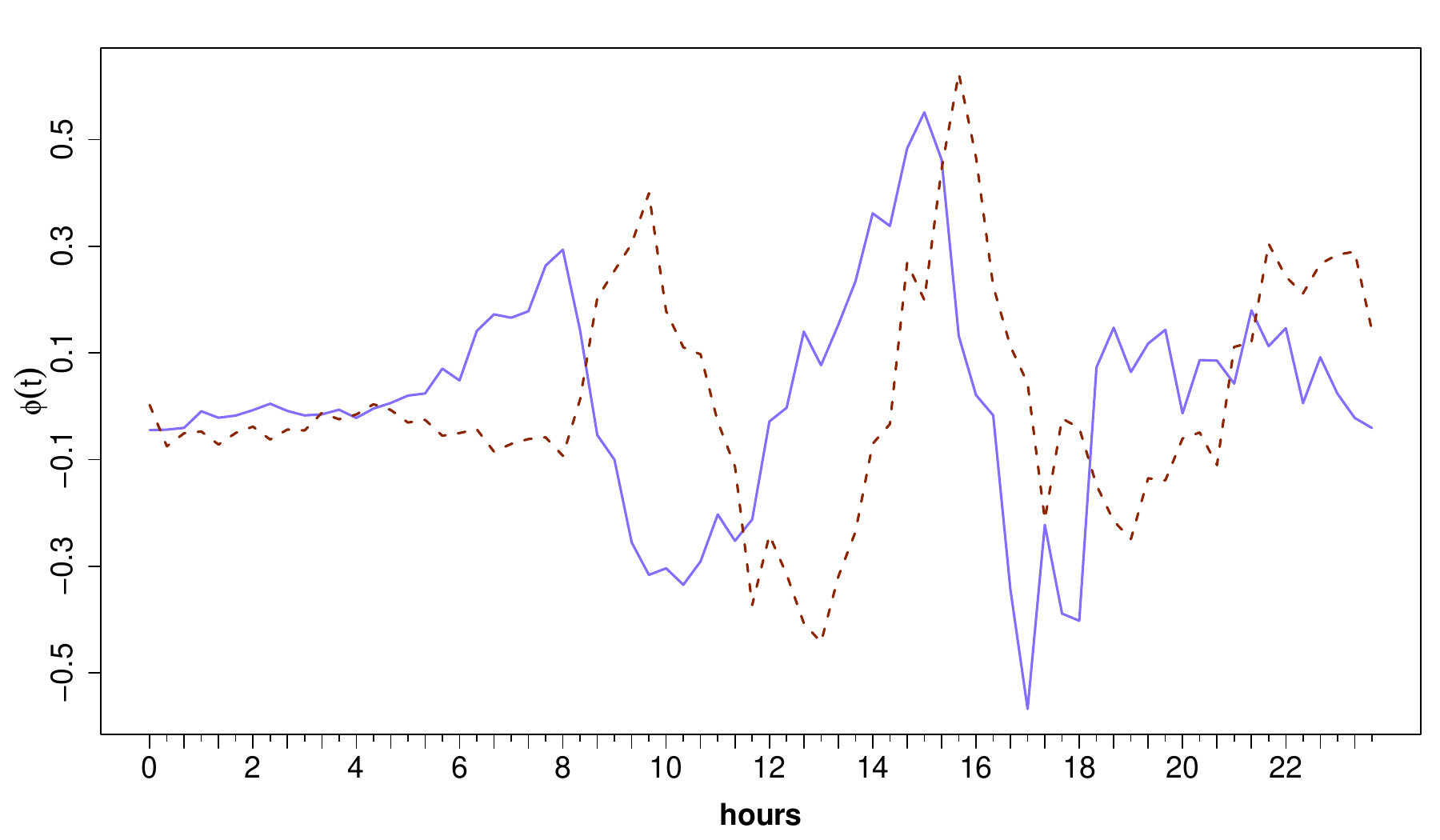}
    \caption{The $5$th eigenfunction (solid line) and the $6$th eigenfunction (dash line) for the bike-sharing data.}
    \label{fig:bike_eigenfunctions_extra}
\end{figure}

This section provides additional figures to demonstrate the dataset and the performance of the proposed method. Figure \ref{fig:the_number_of_rentals_over_time} and \ref{fig:total_rental_time} illustrate the hourly rental profiles over $24$ hours for all days in the training set and the histogram of the total rental times in these days, respectively. Based on the estimated order for the capital bike data, we plot the estimated slope function $\hat{\beta}$ when $\hat{d} = 5$. Furthermore, figure \ref{fig:bike_eigenfunctions} shows the first three estimated eigenfunctions, while \ref{fig:bike_eigenfunctions_extra} illustrates the $6$th and the $7$th estimated eigenfunctions.

\begin{figure}
    \centering
    \includegraphics[scale = 0.5]{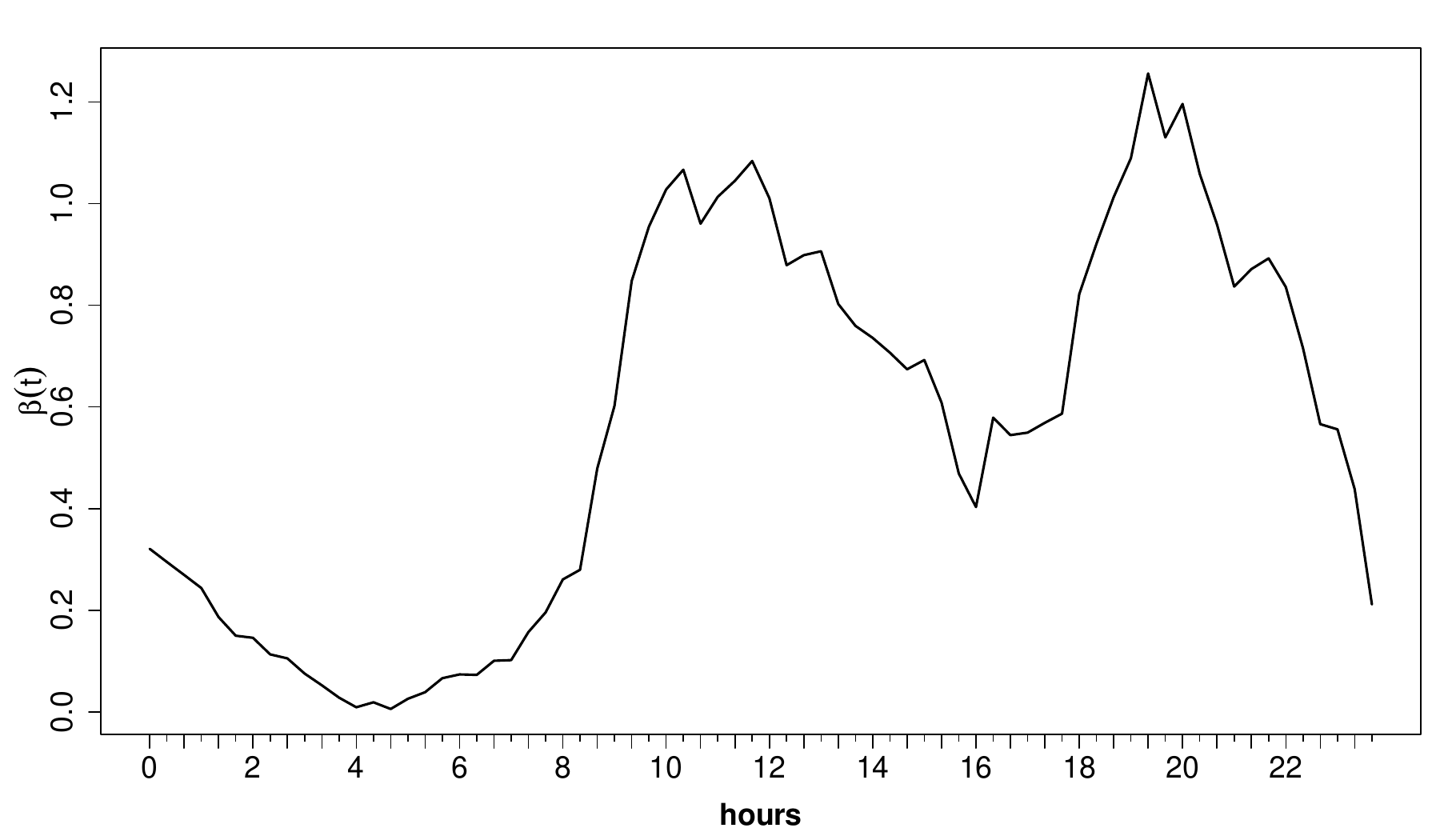}
    \vspace{-1em}
    \caption{The estimated slope function, $\hat{\beta}(t)$, for the bike-sharing dataset, using the first four estimated eigenfunctions.}
    \label{fig:bike_slope_function}
\end{figure}
\clearpage

\clearpage
\bibliographystyle{apalike}
\bibliography{bib.bib}
\end{document}